\documentclass[12pt]{iopart}
\usepackage{graphicx}
\usepackage{lscape}
\usepackage{iopams}
\usepackage{color}
\usepackage{harvard}
\eqnobysec
\begin{document}
\title[Polarizability of cylindrically confined hydrogen atom]{Static and dynamic dipole polarizability of cylindrically confined hydrogen atom}
\author{S.A. Ndengu\'e\footnote{E-mail: sndengue@univ-douala.com}, O. Motapon\footnote{E-mail: omotapon@univ-douala.com}, R.L. Melingui Melono and A. J. Etindele}
\address{Laboratoire de Physique Fondamentale. UFD Math\'ematiques, Informatique Appliqu\'ee et Physique Fondamentale, Universit\'e de Douala, B.P 24157 Douala, Cameroun.}
\begin{abstract}
The non-relativistic static and dynamic dipole polarizabilities of hydrogen atom experiencing a cylindrical confinement are investigated.
Two methods based on B-Splines are used for the computations of the energies and wavefunctions.
The first method is a variational based method while the second one proceeds by a fit of the non-separable Coulomb potential in the product form.
The computed energies compare very well with previous computations.
They converge, as well as the dipole polarizability, to the exact unconfined free atom values.
The fit approach is found to be advantageous, as it helps saving the computational time without a loss of accuracy.
Dynamic polarizabilities have been reported for various dimensions of the confining cylinder.
\end{abstract}
\noindent{\it Keywords: Dipole polarizability, confinement, b-splines, sum of products form.\/}
\noindent\pacs{31.15.ap, 31.15.ag, 31.15.xt}
\submitto{\JPB}
\maketitle

\section{\textbf{Introduction}}
The idea of confined quantum system originates from the pioneering work of \citeasnoun{michels1937} who used an impenetrable cavity to simulate the effect of pressure on the static polarizability of a hydrogen atom.
Under special constraints - spatial restriction under pressure, trapping inside a hollow cage (fullerene cage) or an attractive potential - atoms and molecules undergo numerous changes in their properties (orbitals, energy levels, localization of electrons, polarizability, filling of shells, photon induced ionization and absorption, etc.) compared to the free systems.
In the past two decades, confined atoms have known a rising interest owing to the whole bunch of potential applications: hydrogenic impurities in semiconductors/nanostructures (quantum dots, quantum wells, quantum well wires), atoms emprisoned in zeolite traps, atoms trapped in fullerene cages, etc.
Detailed discussions of these applications are available on several review articles \cite{dolmatov2004,connerade2003,jaskolski1996}.
Recently, two consecutive volumes of {\em Advances in Quantum Chemistry} \cite{sabin2009a,sabin2009b} provided state of the art advances in the study of confined atoms and molecules: 
the principal idea one may consider for the (numerical) attractivity (besides the potential applications) of those studies is the inadequacy of most standard quantum chemistry programs to obtain results for confined systems; that is most of the time, new or modified standard codes and approaches have to be proposed to tackle issues related to confinement.

The initial study on confined atoms, as recalled above, was on the polarizability of the hydrogen experiencing a spherical confinement.
The dipole polarizability of an atom describes the in lowest order the distortion of the electron cloud in the presence of an external electric field.
It is linked to many physical quantities such as the dielectric constant, the refractive index, the ion mobility in gas, the Van der Waals constant, the long range electron-atom interaction potential or even the core potential in model potential calculations.
In the past decades, many works having been devoted to the computations of the polarizability of confined systems and particularly of the simplest one, the hydrogen atom, using a variety of techniques and different types of potential.
We can for instance cite \citeasnoun{montgomery2002}, \citeasnoun{saha2002}, \citeasnoun{laughlin2004}, \citeasnoun{burrows2005}, \citeasnoun{cohen2008}, \citeasnoun{ndengue2008}, \citeasnoun{saha2002} who used hard wall spherical, model endohedral or even model Debye plasma potential to simulate various types of confinement.
The ideal situation of spherical confinement is convenient for analytical or numerical treatment due to the separability of the Schr\"odinger equation.
A less ideal situation would be to consider the atom or more generally a system confined in a cylindrical cavity.
This could happen if one considers an atom or an exciton trapped in a nanotube, a hydrogenic impurity in a quantum well wire \cite{bryant1984,brown1986,brown1987} or a semiconductor \cite{greene1983a,greene1983b} or even any other type of trapping by a cylindrical potential \cite{bastard1981,bastard1982}.
In this situation, except for some special types of potentials \cite{eisenhart48}, the Schr\"odinger equation is no longer completely separable and we therefore have to deal with a two dimensional equation with a Coulomb or any other type of potential.

This paper, about the polarizability of cylindrically confined hydrogen atom, highlights the product form method as an accurate time saving procedure for multidimensional calculations on atomic systems.
After the initial works of \citeasnoun{yurenev06} and \citeasnoun{yurenev08} who studied the energy and stability of the hydrogen atom confined in a perfect cylinder, we first revisit previous calculations of the non-relativistic energies of cylindrically centered confined atom and follow our current interest on confined systems \cite{ndengue2008,motapon2011} in computing the static and dynamic dipole polarizabilities.

The paper is organized as follows: in the second section, we present the theoretical approach used to obtain the energies and wavefunctions.
The third section carries on the procedure to obtain the static and dynamic dipole polarizability with some convergence tests results.
The fourth section on results presents and discusses polarizability computations of the confined system.
We end this article by concluding remarks and possible future works.
In the following, unless explicitly stated, the cylinder will refer to a perfect cylinder, that is a cylinder which has equal diameter and height.

\section{Theoretical approach}

\subsection{Description of the study}

We describe the hydrogen atom experiencing a cylindrical confinement as a one electron atom fixed at the geometry center of the impenetrable wall cylinder of length $z_{max}$ and radius $\rho_{max}$.
The Schr\"odinger equation $H\Psi=\epsilon\Psi$ obeyed by our system then reads:
\begin{equation}
\lbrack-\frac{1}{2\rho}\frac{\partial}{\partial\rho}\left(\rho\frac{\partial}{\partial\rho}\right)-\frac{1}{2\rho^2}\frac{\partial^2}{\partial\phi^2}-\frac{1}{2}\frac{\partial^2}{\partial z^2}-\frac{Z}{\sqrt{z^2+\rho^2}}\rbrack\Psi=\epsilon\Psi
\label{eq:schrodinger}
\end{equation}
with $0\leq\rho\leq \rho_{max}$, $\frac{-z_{max}}{2}\leq z\leq\frac{+z_{max}}{2}$ and $0\leq\phi\leq 2\pi$.
The wavefunction, solution to equation \ref{eq:schrodinger} is written as:
\begin{equation}
\Psi(\rho,z,\phi)=\psi(\rho,z)\frac{e^{im\phi}}{\sqrt{2\pi}}
\label{eq:wavefunction1}
\end{equation}
and satisfies the following boundary conditions:
\begin{equation}
\psi(\rho,z_{max}/2)=\psi(\rho,-z_{max}/2)=\psi(\rho_{max},z)=0
\label{eq:boundaryconditions}
\end{equation}
That is the wavefunction is null on the walls of our cylinder. 
This is equivalent to solving the Schr\"odinger equation with a potential having the Coulomb type inside the cylinder domain, but infinite outside.\\
\newline
\Eref{eq:schrodinger} is solved by expanding the wavefunction in a basis of functions.
Because of their flexibility and their adequacy for confined systems study, we choose a basis of splines (B-Splines) which definition and properties were presented by De Boor \cite{deboor78} and reviewed in spherical atomic physics papers \cite{bachau2001,sapirstein1996}.

\subsection{Variational method}

The non angular part of the wavefunction is expanded as a linear combination of products of functions in $\rho$ and $z$:
\begin{equation}
\psi(\rho,z)=\sum_{i=1}^{N}C_{i}f_{i}(\rho)g_{i}(z)
\label{eq:wavefunction}
\end{equation}
with $N$ being the number of functions and $C_{i}$ the real coefficients of the expansion to be determined.
The functions $f$ and $g$ are B-splines functions:
\begin{equation}
f_{i}(\rho)=B_{\alpha}^{k_{\rho}}(\rho); \\
g_{i}(z)=B_{\beta}^{k_{z}}(z)\\
\label{eq:functionsfandg}
\end{equation}
The $i$ index is expressed as:
\begin{equation}
i=(\alpha-1){\tilde{\eta}_{u}}+\beta; \\ 
\alpha=1, \tilde{\eta}_{\rho}; \\  
\beta=1, \tilde{\eta}_{u}. \\ 
\label{eq:irelations1}
\end{equation}
\begin{equation}
\tilde{\eta}_{\rho}=\eta_{\rho}-1; \\
\tilde{\eta}_{u}=\eta_{u}-2. \\
\end{equation}
The $-1$ and $-2$ in the last relation come from the exclusion of the edge splines, following our boundary conditions.
This happens since the first and last B-Splines are non-zero on the boundary of the domain they span.

When the atom is placed at the center of our cylinder, the system has a $D_{\infty h}$ symmetry \cite{yurenev08} and the confined states are defined by the quantum number $m$, the absolute value of the projection of angular momentum onto the $z$ axis of the cylinder.
Also, the parity of the states with respect to inversion are denoted by $\mathcal{P}$ taking values $+1$ and $-1$, corresponding to $gerade$ and $ungerade$ states.
Taking advantage of those elements we reduce by four (selectively computing $gerade$ or $ungerade$ states) the size of our basis by conveniently defining our wavefunctions \cite{vanne04}.
The $f$ and $g$ functions change to:
\begin{equation}
f_{i}(\rho)=B_{\alpha}^{k_{\rho}}(\rho); \\
g_{i}(z)=B_{\beta}^{k_{z}}(z)+(-1)^{m}\mathcal{P}B_{\eta_{z}+1-\beta}^{k_{z}}(z)\\
\label{eq:functionsfandgbetter}
\end{equation}
now, with
\begin{equation}
\tilde{\eta}_{\rho}=\eta_{\rho}-1; \\
\tilde{\eta}_{u}=(\eta_{u}-2)/2. \\
\end{equation}

The variational transformation of the Schr\"odinger equation \eref{eq:schrodinger} leads to a set of N$\times$N generalized eigenvalue equation on the energies $\varepsilon$ and the coefficients $C_{i}$.
\begin{equation}
Hv=\varepsilon Sv
\label{eq:geneigenvalueeqn}
\end{equation}
where $H$ is the Hamilltonian matrix and $S$ the overlapping matrix. $v$ is the vector of $C_{i}$ coefficients.

The main drawback of this approach (hereafter called 'Method 1') is the high computational time required to converge the calculations, which is due to the non separability of the equation \cite{eisenhart34,eisenhart48}.
As a matter of fact, the time spent to compute the potential double integral increases significantly with the basis size.
This procedure leads to accurate energies and wavefunctions for the ground and excited states with a substantial computational time coming from: $(i)$ the calculation of the Coulomb potential double integral; $(ii)$ the diagonalization of large matrices.
Comparison of results obtained from this approach with previous works \cite{yurenev06,yurenev08} on cylindrical systems and limiting case of the free hydrogen atom will be presented in section \ref{ssec:energyand wavefunctions}.

\subsection{Fit of Coulomb potential}

Two aspects were pointed out as the main sources of important computational time.
While the diagonalization could be resolved by selecting optimized routines, taking advantage of the symmetry and the specificity of the B-Splines basis (leading to band matrices), the calculation of the Coulomb double integrals, less flexible, depends on the performance of the integration procedure.
We used here the Gaussian quadrature, taking about 64$\times$64 points in the $\rho$ and $z$ directions to converge our double integrals.
This could be overcome by applying a method currently used in molecular physics (hereafter called 'Method 2'), that consists in writing the Coulomb potential in the product form \cite{jackle96,jackle98,bowman08}.
Here, the Coulomb potential is expressed as a linear combination of products of functions depending on $\rho$ and $z$.
\begin{equation}
V(\rho,z)=\frac{-Z}{\sqrt{\rho^{2}+z^{2}}}=\sum_{j=1}^{M}D_{j}B_{\gamma}^{kf_{\rho}}(\rho)B_{\zeta}^{kf_{z}}(z)
\label{eq:potentialfit}
\end{equation}
The $D's$ are the coefficients of the expansion and the $B$'s are B-Splines functions of order $kf_{\rho}$ and $kf_{z}$ (generally low - cubic or quartic - in order to reduce the Runge phenomenon \cite{runge1901}).
The $j$, $\gamma$ and $\zeta$ are expressed through relations similar to \eref{eq:irelations1}.
Following the fit of the Coulomb potential according to equation \eref{eq:potentialfit}, the unseparable double integral on the potential which requires $64\times64$ points is now reduced to a value proportional to $64+64$ since the B-Splines functions in equation \eref{eq:potentialfit} have a little support and thus many of the one dimension integrals will cancel.
More, the B-Splines integrals are saved to ease the computation of the term involving the Coulomb integral in the Hamiltonian matrix.

The fit of the Coulomb potential is performed using the \textit{regrid} subroutine of the Dierckx package \cite{dierckx80,dierckx82,dierckx93} from the NETLIB repository.
This is performed from exact rectangulary sampled data in our confined rectangle (cylinder in 2D).
The routine allows a fast and adjustable calculation of the expansion coefficients but still suffers from the Runge phenomenon \cite{runge1901}, that is the oscillation at the boundaries of the interval coming from the polynomial interpolation.
The more drastic problem arises close to the center of the cylinder $(\rho=0,z=0)$ where the accuracy of the results for (at least) the ground state wavefunction and energy relies on the accuracy of the potential, as the principal structure of the wavefunction is localized in that region.
Strictly speaking, there is no recipe to solve exactly this issue but some tricks could help improve the quality of the fit: 
\textit{(i)}   Data sampling increase close to the center of  the cylinder and use of exponential knot sequences so as to increase the number of functions that span the region close to the origin.
\textit{(ii)}  Use of least squares fitting.
\textit{(iii)} Piecewise polynomials number increase in the problematic domain.

The fit using the \textit{regrid} subroutine were performed by an optimization of the B-Splines representation of the Coulomb potential in the $\rho$ and $z$ directions.
Thus, once the exact fitted Coulomb potential points were computed, the routine allowed to find the best B-representation for a certain smoothness parameter.
\Tref{tab:fit-table-1} shows the residual of the fit with increased sampling of data, and the generated basis of quartic spline functions in $\rho$ and $z$, the smoothness selected being $10^{-5}$.
An equally spaced grid sampling has been selected here.
We see, as expected, that the residual of the fit decreases with the increase of sampling data, though not in a regular fashion.
This comes from the use of the \textit{regrid} subroutine which seeks the best compromise between the smoothness of data and the number of B-Splines expansion functions.

\begin{table}
\caption{\label{tab:fit-table-1}Fitted output basis and residual ($\chi^{2}$) from the \textit{regrid} subroutine.
Sampling data are selected on an equally spaced grid.
The smoothness has value of $10^{-5}$.}
\begin{indented}
\item[]
\begin{tabular}{@{}ccccc}
\br
Fit data $\rho$ & Fit data $z$ & Output basis $\rho$ & Output basis $z$ & $\chi^{2}$ \\
\mr
 20 & 40 & 22 & 31 & 5.85(-5) \\
 40 & 92 & 23 & 39 & 2.69(-5) \\
 60 & 102 & 28 & 42 & 2.07(-5) \\
 80 & 102 & 30 & 42 & 1.79(-5) \\
 100 & 102 & 32 & 43 & 1.60(-5) \\
 400 & 800 & 47 & 70 & 2.80(-6)\\
 800 & 1600 & 50 & 84 & 1.40(-6)\\
 1200 & 2400 & 64 & 92 & 9.33(-7)\\
\br
\end{tabular}
\end{indented}
\end{table}

\subsection{\label{ssec:energyand wavefunctions} Energy levels and Wavefunctions}

\begin{figure}
\caption{\label{fig:converge}Convergence of the $1\sigma_{g}$ energy when a linear (straight line) and an exponential (dashed line) sampling of data is used with respect to the number of data.
The boundaries are at $\rho_{max}=20.0$ a.u. and $z_{max}=40.0$ a.u.
Three basis size are presented: N$_{\rho}$=20, 40 and 60. 
The straight line at E=0.5 a.u. is the reference free hydrogen atom ground state energy.}
\includegraphics{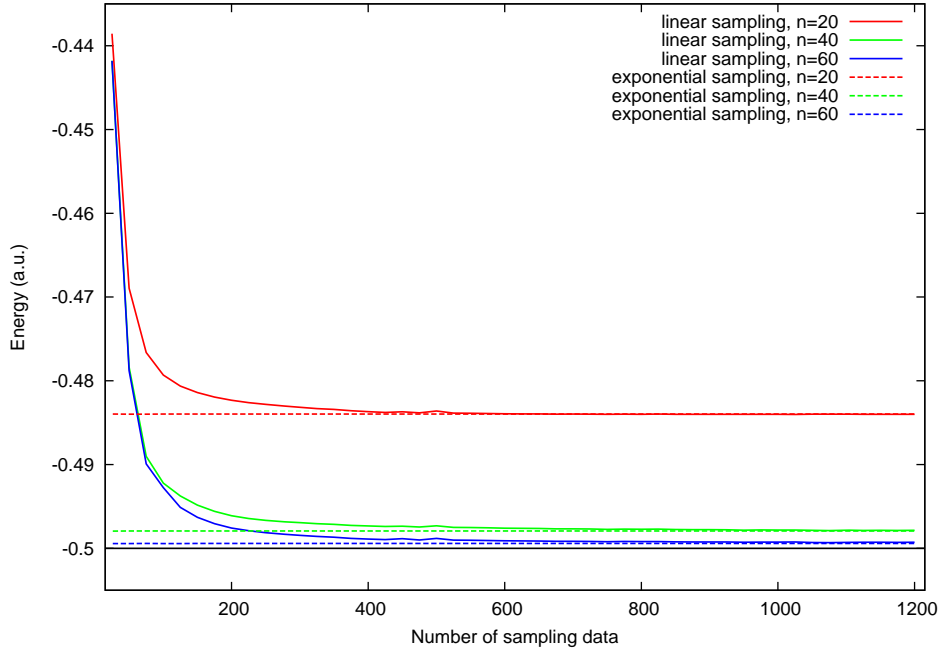}
\end{figure}

The energies and wavefunctions are obtained from the generalized eigenvalue problem of equation \eref{eq:geneigenvalueeqn}, whether we proceed through Method 1 or Method 2.
While Method 1 is a strict variational procedure, where energies converge to numerical exact results with increasing basis, Method 2 is more subtle as convergence depends not only on the increment of the B-Splines basis, but also on the good quality of fit, that is on the type of the sampling performed (linear, exponential, etc) or on the number of sample data used to generate the Coulomb potential, as we pointed out above, mentionning tricks that could help improve the quality of the fit.
This is shown in \fref{fig:converge} where we present the convergence of energies of the ground state  $1\sigma_{g}$ with: increasing number of sampling data, different type of sampling (linear and exponential) and three basis functions sizes (20, 40 and 60 functions in the $\rho$ direction, the double in the $z$ direction).
The results are compared with the exact results for a free hydrogen atom (the straight line at -0.5 a.u.), here considering the infinity to have value $\rho_{max}=20.0$ a.u. and $z_{max}=40.0$ a.u.
As it can be seen on the figure, the energies using an exponential sampling of data converge quickly to their asymptotic value with respect to the basis size.
This feature is true for all the low lying states of our system.
The linear sampling is more helpful when one needs to describe highly excited states which have a long extension on the grid.

We compute the energies of hydrogen atom confined in a perfect cylinder for various confinement values using Methods 1 and 2 using a linear sampling of 1000$\times$2000 data along the $\rho$ and $z$ directions with N$_{\rho}$=50 and N$_{z}$=100 basis functions. 
We compare our results with previous calculations of \citeasnoun{yurenev08} as shown on tables \ref{tab:compare-1}, \ref{tab:compare-2} and \ref{tab:compare-3}.
We can see that our results are of comparable accuracy and agree well (up to five and six significant figures) with those of \citeasnoun{yurenev08}.
This stresses out the accuracy of the fit performed here and its adequacy for the description of the polarizability.
In the same table we report the energies for the same diameter in the case of an infinitely long cylinder.
The values reported here are obtained by considering the infinity for $z_{max}=40.0$ a.u.
For that value of $z_{max}$ (or the corresponding value for $\rho_{max}$), the ground state wavefunction is evanescent and closed to 0, its value at infinity, either in the $z$ or the $\rho$ directions.
This assumption is true for the ground state but also for the lowest excited states that have the greatest contribution to the polarizability.
We see that as the confinement is released along the $z$ direction the energy levels ordering changes and the number of states close to the ground state increases.
This can be seen on \tref{tab:compare-1} where all the 8 lowest energy states are $\sigma$ states whereas $\pi$ and $\delta$ states are inserted in the list in the case of the perfect cylinder.
Also, those eight lowest states in the case of the infinitley long cylinder only have about 2 a.u of difference whereas the two lowest energy states of the corresponding perfect cylinder are separated by more than 4 a.u.

The variational approach with B-Splines functions presented here is stable, unlike the finite difference method of \citeasnoun{yurenev08} as they remarked in their paper.
We are able, as the compared calculation in the free atom case shows (see \tref{tab:polarizability-1}), to go beyond $\rho_{max}=10.0$ a.u.
The only limitation we are facing is on the size of the basis needed to converge results for relatively large cavities: we will then have to deal will large basis leading to an exponential increase in computational time.
Still, for the small cavity dimensions, the variational procedure is numerical exact and the results are accurate at least to six or seven significant figures for the energy and four to five figures for the polarizability.
We cope with those issues by paralellizing our code and using convenient subroutines for large and sparse matrices.

\begin{table}
\caption{\label{tab:compare-1}Comparison of the energy levels obtained with Methods 1 and 2 with the results of \cite{yurenev08}.
The boundaries are at $\rho_{max}=1.0$ a.u and $z_{max}=2.0$ a.u.
The last column reports results of the infinitely long cylinder where the corresponding states are in parenthesis.}
\begin{indented}
\item[]
\begin{tabular}{@{}ccccc}
\br
 State & \cite{yurenev08} & Method 1 & Method 2 & Infinite Cylinder \\
\mr
 1$\sigma_{g}$ &  1.786956 &  1.786914 &  1.786975 & 1.280401 (1$\sigma_{g}$) \\ 
 1$\sigma_{u}$ &  6.192720 &  6.192664 &  6.192664 & 2.516120 (1$\sigma_{u}$) \\ 
 1$\pi_{u}$    &  6.873976 &  6.873849 &  6.873849 & 2.688464 (2$\sigma_{g}$) \\ 
 1$\pi_{g}$    & 10.881392 & 10.881406 & 10.881406 & 2.803335 (2$\sigma_{u}$) \\ 
 2$\sigma_{g}$ & 12.030815 & 12.030945 & 12.031026 & 2.881405 (3$\sigma_{g}$) \\ 
 1$\delta_{g}$ & 12.925836 & 12.925433 & 12.925433 & 3.018616 (3$\sigma_{u}$) \\ 
 3$\sigma_{g}$ & 14.233863 & 14.232317 & 14.232393 & 3.145279 (4$\sigma_{g}$) \\ 
 2$\delta_{u}$ & 16.839909 & 16.838382 & 16.838382 & 3.344140 (4$\sigma_{u}$)\\ 
\br
\end{tabular}
\end{indented}
\end{table}

\begin{table}
\caption{\label{tab:compare-2}Same as \tref{tab:compare-1} with $\rho_{max}=2.0$ a.u and $z_{max}=4.0$ a.u.}
\begin{indented}
\item[]
\begin{tabular}{@{}ccccc}
\br
 State & \cite{yurenev08} & Method 1 & Method 2 & Infinte Cylinder \\
\mr
 1$\sigma_{g}$ & -0.221184 & -0.221201 & -0.221146 & -0.279120 (1$\sigma_{g}$) \\ 
 1$\sigma_{u}$ &  1.126979 &  1.126957 &  1.126957 & 0.422473 (1$\sigma_{u}$) \\ 
 1$\pi_{u}$    &  1.280031 &  1.279982 &  1.279982 & 0.552022 (2$\sigma_{g}$) \\ 
 1$\pi_{g}$    &  2.367238 &  2.367229 &  2.367229 & 0.655296 (2$\sigma_{u}$) \\ 
 2$\sigma_{g}$ &  2.427099 &  2.427165 &  2.427241 & 0.737999 (3$\sigma_{g}$) \\ 
 1$\delta_{g}$ &  2.852432 &  2.852313 &  2.852313 & 0.878874 (3$\sigma_{u}$) \\ 
 3$\sigma_{g}$ &  3.050422 &  3.050048 &  3.050073 & 1.011556 (4$\sigma_{g}$) \\ 
 2$\pi_{u}$    &  3.883131 &  3.883732 &  3.883732 & 1.163751 (1$\pi_{u}$) \\ 
\br
\end{tabular}
\end{indented}
\end{table}

\begin{table}
\caption{\label{tab:compare-3}Same as \tref{tab:compare-1} with $\rho_{max}=4.0$ a.u and $z_{max}=8.0$ a.u.}
\begin{indented}
\item[]
\begin{tabular}{@{}ccccc}
\br
 State & \cite{yurenev08} & Method 1 & Method 2 & Infinite Cylinder \\
\mr
 1$\sigma_{g}$ & -0.489779 & -0.489779 & -0.489756 & -0.491863 (1$\sigma_{g}$) \\ 
 1$\sigma_{u}$ &  0.059391 &  0.059378 &  0.059378 & -0.044154 (1$\sigma_{u}$) \\ 
 1$\pi_{u}$    &  0.088821 &  0.088806 &  0.088806 & 0.032078 (2$\sigma_{g}$) \\ 
 2$\sigma_{g}$ &  0.271533 &  0.271560 &  0.271577 & 0.074807 (1$\pi_{u}$) \\ 
 1$\pi_{g}$    &  0.411472 &  0.411463 &  0.411463 & 0.102282 (2$\sigma_{u}$) \\ 
 1$\delta_{g}$ &  0.519412 &  0.519379 &  0.519379 & 0.132850 (3$\sigma_{g}$) \\ 
 3$\sigma_{g}$ &  0.540286 &  0.540211 &  0.540213 & 0.199615 (3$\sigma_{u}$) \\ 
 2$\pi_{u}$    &  0.781943 &  0.781914 &  0.781914 & 0.240867 (4$\sigma_{g}$) \\ 
\br
\end{tabular}
\end{indented}
\end{table}

\section{Electric dipole polarizability}

The electric dipole polarizability materializes the second order response of the atom in a weak electric field.
From perturbation theory, using the second order energy correction, the polarizability is obtained from the sum over states formula.
An atom cylindrically confined initially in the $n_{0}$ state (where $n_{0}$ includes all the quantum numbers), and experiencing a time independent electric field, possesses a static electric dipole polarizability written as:
\begin{equation}
\alpha_{z}(\omega)=\alpha_{\|}(\omega)=2\sum_{k\neq 0}\frac{|<\Psi_{0}(\rho,z,\phi)|z|\Psi_{k}(\rho,z,\phi)>|^{2}}{E_{k}-E_{0}}
\label{eq:polarizabilityz}
\end{equation}
where  the $n$ is droped to ease the notation.

This component being considered as the parallel component, the orthogonal components along the $x$ and $y$ directions are written as:
\begin{equation}
\alpha_{\bot}=\frac{1}{2}(\alpha_{\bot x}+\alpha_{\bot y})
\label{eq:polarizabilityxy}
\end{equation}
with
\begin{equation}
\alpha_{\bot x}=2\sum_{k\neq 0}\frac{|<\Psi_{0}(\rho,z,\phi)|x|\Psi_{k}(\rho,z,\phi)>|^{2}}{E_{k}-E_{0}}
\label{eq:polarizabilityx}
\end{equation}
and a similar equation for the $y$ component when we replace $x$ by $y$.
The total polarizability is then written as:
\begin{equation}
\alpha=(\alpha_{\|}+2\alpha_{\bot})/3
\label{eq:polarizability}
\end{equation}

In those relations, $x$ and $y$ respectively have values:
$x=\rho cos\phi$ and $y=\rho sin\phi$.

The integration of the angular part of the relations \ref{eq:polarizabilityz} and \ref{eq:polarizabilityxy} leaves the relations with selection rules on orbital states and parity.
For instance, starting with a state of magnetic quantum number $m$, the non zero couplings conditions will be $\Delta m=0$ for the $\alpha_{\|}$ component and $\Delta m=\pm 1$ for the $\alpha_{\bot}$ component.

If instead the atom is exposed to a radiation field of frequency $\omega$, the frequency dependent dipole polarizability will then write for the parallel component:
\begin{equation}
\alpha_{z}=\alpha_{\|}=2\sum_{k\neq 0}\frac{|<\Psi_{0}(\rho,z,\phi)|z|\Psi_{k}(\rho,z,\phi)>|^{2}}{(E_{k}-E_{0})^{2}-\omega^{2}}(E_{k}-E_{0})
\label{eq:polarizabilitydynamic}
\end{equation}

The calculation of the polarizability using the sum-over-states (SOS) method requires accurate evaluation of energies and wavefunctions of the system. 
The polarizability of the ground state is sensitive to the accuracy of the first few lowest states as the major contribution comes from the overlapping of the ground state with the first few excited states, but the accuracy of its calculation depends on how precise the energies and wavefunctions of the ground state are evaluated.
This assertion is visible on \tref{tab:polarizability-1} where we computed the static polarizability of the ground state using a linear and an exponential sampling of data points with increasing splines basis size.
We used 1000$\times$2000 sampling points along the $\rho$ and $z$ directions.
The exponential sampling which leads to more accurate results for the ground state energies and wavefuctions than a linear sampling with the same number of points has also a better polarizability convergence with increasing basis size.

\begin{table}
\caption{\label{tab:polarizability-1}Comparison of the ground state static dipole polarizability with increasing linear and exponential data points sampling.
The boundaries are at $\rho_{max}=20.0$ a.u. and $z_{max}=40.0$ a.u.
The 1$^{st}$ column is the number of B-Spline functions in the $\rho$ direction. The value in the $z$ being twice that value.
The 2$^{nd}$ to 5$^{th}$ columns are the ground state energies and polarizabilities with linear and exponential sampling respectively.}
\begin{indented}
\item[]
\begin{tabular}{@{}ccccc}
\br
 Basis size & E$_{lin}$(1$\sigma_{g}$) & E$_{exp}$(1$\sigma_{g}$) & $\alpha_{lin}$(1$\sigma_{g}$) & $\alpha_{exp}$(1$\sigma_{g}$) \\
\mr
 20 & -0.4840 & -0.4840 & 5.1611 & 5.2052 \\
 40 & -0.4978 & -0.4979 & 4.5691 & 4.5655 \\
 60 & -0.4993 & -0.4994 & 4.5230 & 4.5180 \\
 80 & -0.4996 & -0.4998 & 4.5118 & 4.5071 \\
100 & -0.4997 & -0.4999 & 4.5078 & 4.5034 \\
120 & -0.4998 & -0.4999 & 4.5060 & 4.5018 \\
 $\infty$ & -0.5000 & -0.5000 & 4.5000 & 4.5000 \\
\br
\end{tabular}
\end{indented}
\end{table}

\section{Results and discussions}

The static dipole polarizability of the hydrogen atom confined in the cylinder follows the same trends as what is observed in the case of the hard wall spherical confinement \cite{montgomery2002,laughlin2004,burrows2005,cohen2008}, that is the polarizability reduces with the confinement radius as can be seen on \tref{tab:compare-spherical}.
However cylindrically confined polarizability is generally greater than the spherically confined value.
This may roughly be justified from the view that the volume accessible for a particular confinement value is greater for the cylinder than for the sphere and thus the electron has more latitude to move away from the nucleus in the cylinder than in the sphere.

The parallel and orthogonal components of the cylindrical polarizability are also shown on the table.
As expected, all components but one of the polarizability are of the same order, whether we consider the perfect or infinitely long confined cylinder or even the spherically confined cylinder.
Only is the parallel component of the infinitely long cylinder that is 4 to ten times higher than the coresponding component for hard confinement (1 and 2 a.u.).
This is simple to understand when we consider the available space to the electron and as expected it reduces as the confinement is released.

\begin{table}
\caption{\label{tab:compare-spherical}Ground state cylindrically and spherically confined static dipole polarizability for various confinement radii.
The 1$^{st}$ column is the confinement radius.
The 2$^{nd}$ and 7$^{th}$ columns are respectively the parallel (equation \eref{eq:polarizabilityz}), orthogonal (equation \eref{eq:polarizabilityxy}) and total (equation \eref{eq:polarizability}) cylindrically confined static dipole polarizability for the perfect and infinitely long cylinder.
The 8$^{th}$ column is the spherically confined polarizability summarized from various references \cite{montgomery2002,laughlin2004,burrows2005,cohen2008}.}
\begin{indented}
\item[]
\begin{tabular}{@{}cccccccc}
\br
 Radius & $\alpha_{\| per}$ & $\alpha_{\bot per}$ & $\alpha_{cyl|per}$ & $\alpha_{\| inf}$ & $\alpha_{\bot inf}$ & $\alpha_{cyl|inf}$ & $\alpha_{sphe}$ \\
\mr
        1 & 0.0506 & 0.0379 & 0.0421 & 0.5130 & 0.0404 & 0.1980 & 0.0288 \\
        2 & 0.5410 & 0.4373 & 0.4719 & 1.6449 & 0.4658 & 0.8588 & 0.3426 \\
        4 & 2.9636 & 2.7205 & 2.8015 & 3.7736 & 2.7883 & 3.1168 & 2.3780 \\
        6 & 4.2869 & 4.2075 & 4.2340 & 4.4210 & 4.2216 & 4.2881 & 4.0581 \\
        8 & 4.4863 & 4.4767 & 4.4799 & 4.4945 & 4.4760  & 4.4821 & 4.4540 \\
       10 & 4.4994 & 4.4987 & 4.4989 & 4.4992 & 4.4985 & 4.4987 & 4.4968 \\
 $\infty$ & 4.5000 & 4.5000 & 4.5000 & 4.5000 & 4.5000 & 4.5000 & 4.5000 \\
\br
\end{tabular}
\end{indented}
\end{table}

\begin{figure}
\caption{\label{fig:polar-multi1}Ground state dynamic dipole polarizability of perfect (red) and infinitely long (black) cylinder for $\rho_{max}$=1 a.u.
We use a frequency step of 0.02 a.u.
(a) is for the parallel component, (b) the orthogonal component and (c) is the total dynamic polarizability.}
\includegraphics{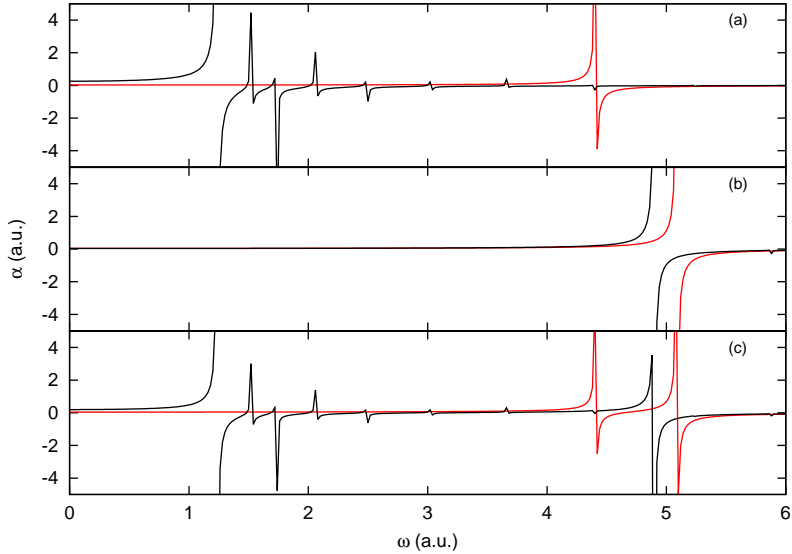}
\end{figure}

\begin{figure}
\caption{\label{fig:polar-multi2}Same as \fref{fig:polar-multi1} with $\rho_{max}$=2 a.u.}
\includegraphics{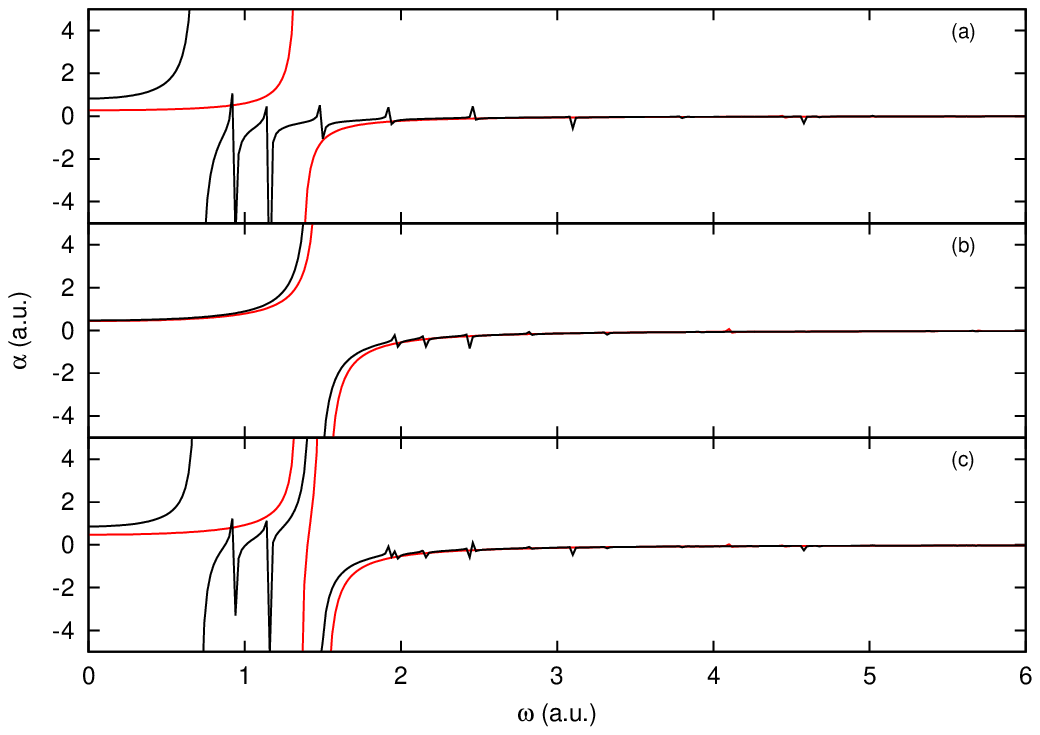}
\end{figure}

\begin{figure}
\caption{\label{fig:polar-multi4}Same as \fref{fig:polar-multi1} with $\rho_{max}$=4 a.u. and a frequency step of 0.002 a.u.}
\includegraphics{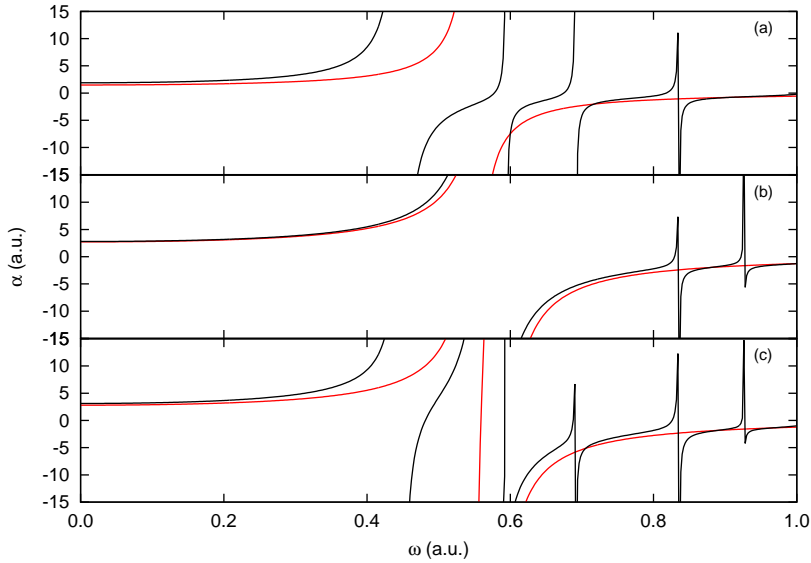}
\end{figure}

\begin{figure}
\caption{\label{fig:polar-multi6}Same as \fref{fig:polar-multi4} with $\rho_{max}$=6 a.u.}
\includegraphics{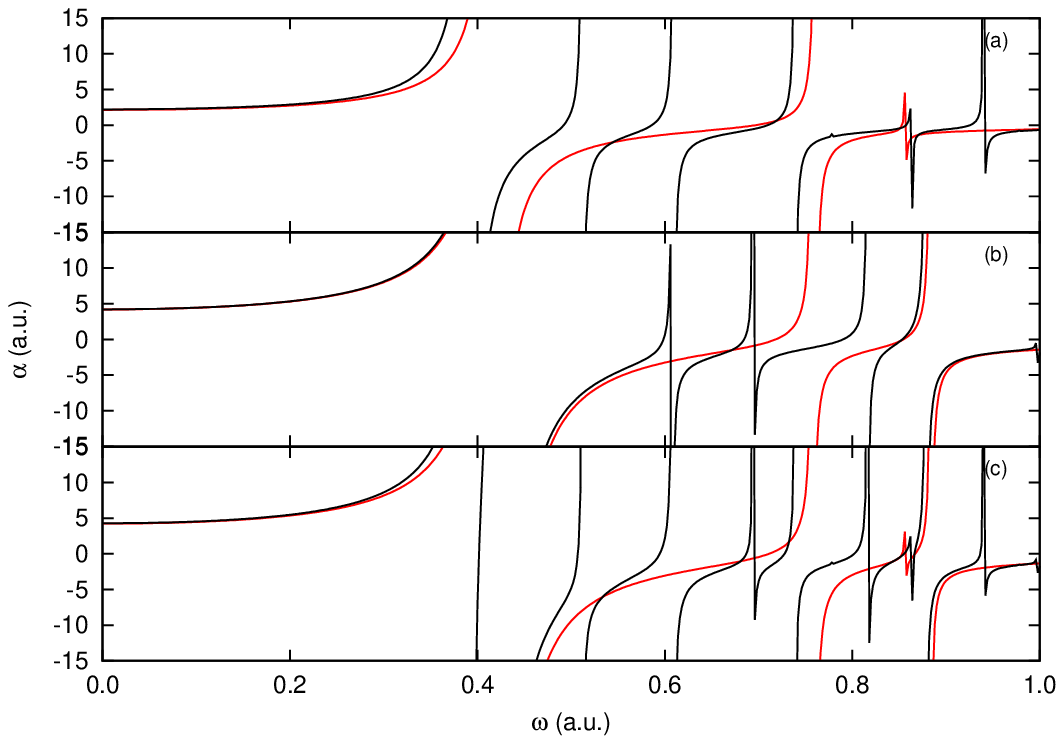}
\end{figure}

Figures \ref{fig:polar-multi1} to \ref{fig:polar-multi6} present the ground state dynamic dipole polarizability of the perfect and infinitely long cylinder for various confinement radii.
The jumps (singularities) appearing on the figures identify frequencies for which the denominators of equations \eref{eq:polarizabilityz} and \eref{eq:polarizabilityxy} cancel.
It is important to stress that those frequencies are not the same as is the case for a spherical confinement because of the non-degeneracy of the states in $m$ and the selection rules mentionned in the preceding section.
Hence the singularities identify either $\Delta m=0$ or $\Delta m=+1$ (starting from a $\sigma$ state) couplings.
It appears on these figures that the position of the first singularity decreases as the confinement is released.
Optimally it should converge to the 0.375 a.u. value of the $1s\rightarrow2p$ of the free hydrogen atom which will correspond to the $1s\sigma\rightarrow2p\sigma$ transition for the hydrogen atom in a large cylinder.
This transition appears on figures \ref{fig:polar-multi1} to \ref{fig:polar-multi6} (panels (a) and (c)) as the very first singularity.

For small confinement values (1 and 2 a.u.), the system presents only one or two singularities for the perfect cylinder in the frequency range of 0 to 6 a.u whereas the infinitely long cylinder presents many more structures as could have been expected from the tables \ref{tab:compare-1} and \ref{tab:compare-2} where many states close to the ground state are observed in the infinitely long cylinder compared to the perfect cylinder.
On a smaller frequency range (0 to 1 a.u.) for slightly released radial confinement (\fref{fig:polar-multi4} and \fref{fig:polar-multi6}) the infinitely long cylinder cases still exhibit more structures than the perfect cylinder.
The singularities on all the figures are always shifted to the left for the infinitely long cylinder.

\section{Conclusion}

A B-Splines based variational method using the sum of products for the representation of the Coulomb potential has been used to compute the polarizability of the hydrogen atom in a cylindrical potential.
The sum of product method is accurate to reproduce the energies of the hydrogen atom confined in a cylinder when compared with previous calculations \cite{yurenev06,yurenev08} and the free atom limit.
The method presented here, on the contrary of \cite{yurenev06} do not suffer from any instabilities in the calculations with increasing cylinder radius or height, though the computation time will increase while seeking convergence with a larger spline basis.

The sum of products representation of the Coulomb potential is convenient for this type of study.
Indeed, it is possible to study the off-center and even off-axis atom in the cylinder when the adequate fit of the modified Coulomb potential is performed.
This will be the content of our future works where a special attention would be drawn to the study of impurities in cylindrical quantum well wire within the framework of single band effective-mass approximation \cite{arunachalam2012}.

\end{document}